\documentclass[pra, amsmath, amssymb, superscriptaddress,reprint]{revtex4-1}

\usepackage{pstricks}
\usepackage{color}
\usepackage{amsmath,amssymb}
\usepackage{epsfig}
\usepackage{graphicx}
\usepackage{dcolumn}
\usepackage{bbold}
\usepackage{natbib}
\usepackage{datetime}
\usepackage{textcomp}
\usepackage{bm}
\usepackage{color}
\usepackage{amsfonts}
\usepackage{blindtext}
\usepackage[utf8]{inputenc}
\usepackage{float}
\usepackage{physics}
\usepackage{braket}
\usepackage{docmute}
\usepackage{ulem}

\usepackage{upgreek}

\begin{document}

\title{Time-dependent optimized coupled-cluster
method with doubles and perturbative triples [TD-OCCD(T)] for first principles simulation of multielectron dynamics}
\author{Himadri Pathak}
\email{hmdrpthk@gmail.com}
\affiliation{Department of Nuclear Engineering and Management, School of Engineering, The University of Tokyo, 7-3-1 Hongo, Bunkyo-ku, Tokyo 113-8656, Japan}
\author{T. Sato}
\email{sato@atto.t.u-tokyo.ac.jp}
\affiliation{Department of Nuclear Engineering and Management, School of Engineering, The University of Tokyo, 7-3-1 Hongo, Bunkyo-ku, Tokyo 113-8656, Japan}
\affiliation{Photon Science Center, School of Engineering, University of Tokyo, 7-3-1 Hongo, Bunkyo-ku, Tokyo 113-8656, Japan}
\affiliation{Research Institute for Photon Science and Laser Technology, University of Tokyo, 7-3-1 Hongo, Bunkyo-ku, Tokyo 113-0033, Japan}
\author{K. L. Ishikawa}
\email{ishiken@n.t.u-tokyo.ac.jp}
\affiliation{Department of Nuclear Engineering and Management, School of Engineering, The University of Tokyo, 7-3-1 Hongo, Bunkyo-ku, Tokyo 113-8656, Japan}
\affiliation{Photon Science Center, School of Engineering, University of Tokyo, 7-3-1 Hongo, Bunkyo-ku, Tokyo 113-8656, Japan}
\affiliation{Research Institute for Photon Science and Laser Technology, University of Tokyo, 7-3-1 Hongo, Bunkyo-ku, Tokyo 113-0033, Japan}
\begin{abstract}
We report the formulation of a new, cost-effective approximation method in the time-dependent optimized coupled-cluster (TD-OCC) framework [T. Sato {\it et al.,} J. Chem. Phys. 148, 051101 (2018)] for {\color{black} first-principles} simulations of multielectron dynamics in an intense laser field. The method, designated as TD-OCCD(T), is a time-dependent, orbital-optimized extension of the ``gold-standard'' CCSD(T) method in the ground-state electronic structure theory. The equations of motion for the orbital functions and the coupled-cluster amplitudes are derived based on the real-valued time-dependent variational principle using the {\color{black} fourth-order} Lagrangian. 
The TD-OCCD(T) is size extensive and gauge invariant, and scales as $O(N^7)$ with respect to the number of active {\color{black} orbitals} $N$.
The pilot application of the TD-OCCD(T) method to the strong-field ionization and high-order harmonic generation from a Kr atom is reported in comparison with the results of the previously developed  method{\color{black}s,} such as the time-dependent complete-active-space self-consistent field (TD-CASSCF), TD-OCC with double and triple excitations (TD-OCCDT), TD-OCC with double excitations (TD-OCCD), and the time-dependent Hartree-Fock (TDHF) methods.
  
\end{abstract}
\maketitle
\section{Introduction}
Recent years witnessed unprecedented progress in laser technologies, which made it possible to observe the motions of electrons at the attosecond time scale (\cite{itatani2004, corkum2007attosecond, krausz2009attosecond, baker2006probing}). 
On the other hand, various theoretical and numerical methods have been {\color{black} developed} for interpreting, understanding{\color{black},} and {\color{black} predicting} the experiments.

The multi-configuration time-dependent Hartree-Fock (MCTDHF) method \cite{caillat2005correlated, kato2004time, nest2005multiconfiguration, haxton2011multiconfiguration, hochstuhl2011two} {\color{black}, and the} time-dependent complete-active-space self-consistent-field (TD-CASSCF) method \cite{sato2013time, sato2016time} are the most rigorous approaches to solve  time-{\color{black}dependent} Schr{\"o}dinger equation (TDSE) of many-electron systems, 
where the wavefunction is given by the full configuration interaction (FCI) expansion,
\begin{eqnarray}
\Psi(t)=\sum_{\bm I}C_I(t)\Phi_{\bm I}(t), 
\end{eqnarray}
with both CI coefficients $\{C_{\bm I}(t)\}$ and orbital functions $\{\psi_p(t)\}$ constituting Slater determinants $\{\Phi_{\bm I}(t)\}$ are propagated in time according to the time-dependent variational principle (TDVP).
The TD-CASSCF method broadens the applicability of the MCTDHF method by flexibly 
classifying the orbital subspace into frozen-core, dynamical-core, and active.
Unfortunately, the factorial computational scaling impedes large-scale applications.
There are reports of various affordable size-inextensive methods \cite{miyagi2013time, miyagi2014time, haxton2015two, sato2015time} developed by limiting the CI expansion of the wavefunction.
Alternatively, the size-extensive coupled-cluster method, which relies on an exponential wavefunction, is a superior choice to address these problems with a polynomial cost-scaling \cite{kummel:2003, shavitt:2009}. 
We have developed an explicitly time-dependent coupled-cluster method considering 
optimized orthonormal orbitals within the flexibly chosen active space, called the time-dependent optimized coupled-cluster (TD-OCC) method, \cite{sato2018communication} including double (TD-OCCD) and double and triple excitation amplitudes (TD-OCCDT).
Our method is a time-dependent formulation of the stationary optimized coupled-cluster method \cite{scuseria1987optimization, sherrill1998energies, krylov1998size}.
Kvaal \cite{kvaal2012ab} also developed an orbital adaptive time-dependent coupled-cluster (OATDCC) method
using biorthogonal orbitals.
We take note of {\color{black} a} few reports on the time-dependent coupled-cluster method{\color{black}s} \cite{huber2011explicitly, pigg2012time, nascimento2016linear},
using time-independent orbitals, and their interpretation \cite{pedersen2019symplectic,  pedersen2020interpretation}, including the very initial attempts \cite{schonhammer1978schonhammer, hoodbhoy1978time, hoodbhoy1979time}. 

The TD-OCCDT scales as $O(N^8)$ ($N$= the number of active orbitals), not ideally suited for applications
to larger chemical systems.
Therefore, we have developed a few-low cost method{\color{black}s} in the TD-OCC framework
\cite{pathak2020timedependent, pathak2020timemp2, pathak2020study, pathak2021time}.
We find triple excitations are necessary, including perfect optimization of the orbitals. Therefore, we are interested in developing affordable TD-OCC methods retaining a part of the triples. 
The most popular coupled-cluster method that treats the triple excitation amplitudes approximately is called CCSD(T) \cite{raghavachari1989fifth, watts1993coupled}.
Bozkaya {\it et al,} \cite{bozkaya2012symmetric} included various {\color{black} symmetric and asymmetric} triple excitation
corrections to their optimized double (OD) method. 

In this communication, we report the formulation and implementation of the CCSD(T) method in the time-dependent optimized coupled-cluster framework, TD-OCCD(T).
Following our previous works \cite{sato2018communication, pathak2020timedependent, pathak2020timemp2, pathak2021time}, we exclude single excitation amplitudes but optimize the orbitals according to {\color{black} time-dependent variational principle} (TDVP).
As the first application of this method, we study electron dynamics in Kr using intense near-infrared laser fields. 

\section{Method}\label{sec2}
The second quantization representation of the Hamiltonian{\color{black},} including the laser field{\color{black},} is as follows, 
\begin{eqnarray}\label{eq:ham2q}
\hat{H}
&=& h^\mu_\nu(t) \hat{c}^\dagger_\mu\hat{c}_\nu +
 \frac{1}{2}u^{\mu\gamma}_{\nu\lambda} \hat{c}^\dagger_\mu\hat{c}^\dagger_\gamma\hat{c}_\lambda\hat{c}_\nu
\label{eq:ham2q-bis}
\end{eqnarray}
where $\hat{c}^\dagger_\mu$ ($\hat{c}_\mu$) represents a creation
(annihilation) operator in a complete,
orthonormal set
{\color{black} of $2n_{\rm bas}$ time-dependent spin-orbitals $\{\psi_\mu(t)\}$.
$n_{\rm bas}$ is the number of basis functions used for expanding the spatial part of  $\psi_\mu$}, which, in the present real-space implementation, corresponds to the number of grid points, and 
\begin{eqnarray}
h^\mu_\nu(t) = \int dx_1 \psi^*_\mu(x_1) [h_0+V_{ext}] \psi_\nu(x_1),\\
 u^{\mu\gamma}_{\nu\lambda} = \int\int dx_1dx_2
 \frac{\psi^*_\mu(x_1)\psi^*_\gamma(x_2)\psi_\nu(x_1)\psi_\lambda(x_2)}{|\bm{r}_1-\bm{r}_2|},
\end{eqnarray}
where $x_i=(\bm{r}_i,\sigma_i)$ represents a composite spatial-spin coordinate.
$h_0$ is the field free one-electronic Hamiltonian and $V_{ext}=A(t)p_z$ in the velocity gauge,
$A(t)=-\int^t E(t^\prime) dt^\prime$ is
the vector potential, with $E(t)$ being the laser electric field
linearly polarized along the $z$ axis.

The complete set of $2n_{\rm bas}$ spin-orbitals (labeled with
$\mu,\nu,\gamma,\lambda$) is divided into $n_{\rm occ}$ {\it occupied} ($o,p,q,r,s$) and
$2n_{\rm bas}-n_{\rm occ}$ {\it virtual} spin-orbitals. 
The coupled-cluster (or CI) wavefunction is constructed only with occupied spin-orbitals{\color{black},} which are time-dependent in general, and virtual spin-orbitals form the orthogonal complement of the occupied spin-orbital space. 
The occupied spin-orbitals
are classified into $n_{\rm core}$ {\it core} spin-orbitals{\color{black},} 
which are occupied in the reference $\Phi$ and kept uncorrelated, and
$N=n_{\rm occ}-n_{\rm core}$ {\it active} spin-orbitals ($t,u,v,w$) among which the active electrons are correlated. The active
spin-orbitals are further split into those in the {\it hole} space
($i,j,k,l$) and the {\it particle} space ($a,b,c,d$), which are defined as those occupied and
unoccupied, respectively, in the reference $\Phi$.
The core spin-orbitals can further be split into {\it frozen-core} space ($i^{\prime\prime},j^{\prime\prime}$),
fixed in time
and the {\it dynamical-core} space ($i^\prime,j^\prime$), propagated in time \cite{sato2013time} (See.
Fig.~1 in ~\cite{sato2018communication} for a pictorial illustration).

The real action formulation of the TDVP with orthonormal
orbitals is our guiding principle, \cite{sato2018communication}
\begin{eqnarray}
S &=& \label{eq:action}
\operatorname{Re}\int_{t_0}^{t_1} Ldt = \frac{1}{2} \int_{t_0}^{t_1}\left( L + L^*\right) dt,\\
L &=&\label{eq:Lag}
 \langle\Phi|(1+\hat{\Lambda})e^{-\hat{T}}(\hat{H}-i\frac{\partial}{\partial
 t})e^{\hat{T}}|\Phi\rangle,\\
\hat{T}&=&\hat{T}_2+\hat{T}_3\cdots=\tau^{ab}_{ij}\hat{E}^{ab}_{ij}+\tau^{abc}_{ijk}\hat{E}^{abc}_{ijk}\cdots, \\
\hat{\Lambda}&=&\hat{\Lambda}_2+\hat{\Lambda}_3\cdots=\lambda_{ab}^{ij}\hat{E}_{ab}^{ij}+\lambda_{abc}^{ijk}\hat{E}_{abc}^{ijk}\cdots,
\end{eqnarray}
where
$\tau^{ab\cdots}_{ij\cdots}$ ($\lambda_{ab\cdots}^{ij\cdots}$) are (de{\color{black}-})excitation amplitudes{\color{black}, and $\hat E_{ij\cdots}^{ab\cdots}=\hat c^{\dagger}_a\hat c^{\dagger}_b \cdots \hat c_j\hat c_i$.}
The stationary conditions,  $\delta S=0$, 
with respect to the variation of the parameters of the wavefunction ($\delta\tau^{ab\cdots}_{ij\cdots}$,
$\delta\lambda_{ab\cdots}^{ij\cdots}$, and $\delta\psi_\mu$)
gives
us the corresponding equations of motions (EOMs), 
$\delta\psi_\mu$ is orthonormality-conserving
orbital variation. 

For deriving the TD-OCCD(T) method,
we first construct a fourth-order Lagrangian defined in \cite{pathak2021time}. 
We make a further approximation to the Lagrangian and write separating it into two parts,
\begin{subequations}
\begin{eqnarray}
L_{\rm CCD(T)}^{(4)}&=&L_0+\langle\Phi|(1+\hat \Lambda_2)[(\bar f+\hat v) e^{\hat T_2}]_c|\Phi\rangle
-i\lambda_{ab}^{ij}\dot \tau_{ij}^{ab}\label{eq:L_ccd(t)_double} \\
&+& \langle\Phi|\hat \Lambda_2 [(\bar f+\hat v)\hat T_3]_c|\Phi\rangle+\langle \Phi|\hat \Lambda_3 (\bar f\hat T_3)_c|\Phi\rangle+\langle \Phi|\hat \Lambda_3(\hat v\hat T_2)_c|\Phi\rangle-i\lambda_{abc}^{ijk}\dot
\tau_{ijk}^{abc}\label{eq:L_ccd(t)_triple},
\end{eqnarray}
\end{subequations}
where $\bar{f}=\hat{f}-i\hat{X}$, 
$\hat{f} = (h^p_q + v^{pj}_{qj}) \{\hat{E}^p_q\}$,
$\hat{v}=v^{pr}_{qs}\{\hat{E}^{pr}_{qs}\}/4$, 
and $v^{pr}_{qs} = u^{pr}_{qs}-u^{pr}_{sq}$, $\hat{X}=X^\mu_\nu\hat{E}^\mu_\nu$, and
$X^\mu_\nu=\langle\psi_\mu|\dot{\psi}_\nu\rangle$ is anti-Hermitian.
The double amplitudes are obtained  
by making $L^{(4)}_{\rm CCD(T)}$ of Eq.~(\ref{eq:L_ccd(t)_double}) stationary
with respect to $\delta S/\delta \lambda^{ij}_{ab}(t) = 0$, 
$\delta S/\delta \tau^{ab}_{ij}(t)= 0$,
the triples by making Eq.~(\ref{eq:L_ccd(t)_triple}) stationary with respect
to
$\delta S/\delta \lambda^{ijk}_{abc}(t) = 0$, and $\delta S/\delta \tau^{abc}_{ijk}(t) = 0$,
%
\begin{eqnarray}
i\dot{\tau}^{ab}_{ij}
&=&\label{eq:td-occd_lint3_t2}
v_{ij}^{ab}-p(ij) \bar{f}_j^k\tau_{ik}^{ab}+p(ab) \bar{f}_c^a \tau_{ij}^{cb} \nonumber \\
&+&\frac{1}{2} v_{cd}^{ab}\tau_{ij}^{cd}
+\frac{1}{2} v_{ij}^{kl} \tau_{kl}^{ab}+p(ij)p(ab)
v_{ic}^{ak} \tau_{kj}^{cb} \nonumber \\ 
&-&\frac{1}{2}p(ij) \tau_{ik}^{ab} \tau_{jl}^{cd} v_{cd}^{kl}
+\frac{1}{2}p(ab) \tau_{ij}^{bc} \tau_{kl}^{ad} v_{cd}^{kl} \nonumber \\
&+&\frac{1}{4} \tau_{kl}^{ab} \tau_{ij}^{cd} v_{cd}^{kl}
+\frac{1}{2}p(ij)p(ab) \tau_{il}^{bc} \tau_{jk}^{ad} v_{cd}^{kl}
\end{eqnarray}

\begin{eqnarray}
-i\dot{\lambda}^{ij}_{ab}
&=&\label{eq:td-occd_l2}
v_{ab}^{ij}-p(ij) \bar{f}_k^i
\lambda_{ab}^{kj}+p(ab) \bar{f}_a^c\lambda_{cb}^{ij} \nonumber \\
&+&\frac{1}{2} v_{ab}^{cd}\lambda_{cd}^{ij}
+\frac{1}{2} v_{kl}^{ij}\lambda_{ab}^{kl}+p(ij)p(ab)
v_{kb}^{cj}\lambda_{ac}^{ik} \nonumber \\ 
&-&\frac{1}{2}p(ij) \lambda_{cd}^{ik} \tau^{cd}_{kl} v_{ab}^{jl}
+\frac{1}{2}p(ab) \lambda_{bc}^{kl} \tau^{cd}_{kl} v_{ad}^{ij} \nonumber \\
&+&\frac{1}{4} \lambda_{ab}^{kl} \tau_{kl}^{cd} v_{cd}^{ij}
+\frac{1}{2}p(ij)p(ab) \lambda_{ac}^{jk} \tau_{kl}^{cd} v_{bd}^{il} \nonumber \\
&-&\frac{1}{2}p(ij) \lambda_{ab}^{ik} \tau_{kl}^{cd} v_{cd}^{jl}\nonumber\\
&+&\frac{1}{2}p(ab) \lambda_{bc}^{ij} \tau_{kl}^{cd} v_{ad}^{kl}+\frac{1}{4} \lambda_{cd}^{ij} \tau_{kl}^{cd} v_{ab}^{kl}
\end{eqnarray}

\begin{eqnarray}
i\dot{\tau}^{abc}_{ijk}
&=&\label{eq:td-occd_lint3_t3}
p(k/ij)p(a/bc)v_{dk}^{bc}t_{ij}^{ad}-p(i/jk)p(c/ab)v_{jk}^{lc}t_{il}^{ab}\nonumber \\
&-&p(k/ij)\bar f_k^l\tau_{ijl}^{abc}+p(c/ab)\bar f_d^c\tau_{ijk}^{abd}, 
\end{eqnarray}

\begin{eqnarray}
-i\dot{\lambda}^{ijk}_{abc}
&=&\label{eq:td-occd_l2}
p(k/ij)p(a/bc)v_{bc}^{dk}\lambda_{ad}^{ij}-p(c/ab)p(i/jk)v_{lc}^{jk}\lambda_{ab}^{ij} \nonumber \\
&+&p(c/ab)\bar f_c^d\lambda_{abd}^{ijk}-p(k/ij)\bar f_l^k\lambda_{abc}^{ijl}\nonumber\\
&+&p(i/jk)p(a/bc)\bar f_a^i\lambda_{bc}^{jk},
\end{eqnarray}

where $p(\mu\nu)$ and $p(\mu|\nu\gamma)$ are the permutation operators; ${\color{black}p(\mu\nu)}A_{\mu\nu}=A_{\mu\nu}-A_{\nu\mu}$, and $p(\mu/\nu\gamma)=1-p(\mu\nu)-p(\mu\gamma)$.

The EOM for the orbitals
can be written down in the following form \cite{sato2016time}, 
\begin{eqnarray}\label{eq:eom_orb}
i|\dot{\psi_p}\rangle &=&
(\hat{1}-\hat{P})
\hat{F}|\psi_p\rangle + i|\psi_q\rangle X^q_p,
\end{eqnarray}
where $\hat{1} = \sum_\mu|\psi_\mu\rangle\langle\psi_\mu|$ is the identity operator within the space spanned by the given basis, 
$\hat{P}=\sum_q|\psi_q\rangle\langle\psi_q|$ is the projector onto the occupied spin-orbital space, and
\begin{eqnarray}
\hat{F}|\psi_p\rangle &=& \label{eq:gfockoperator}
\hat{h} |\psi_p\rangle +  \hat{W}^r_s|\psi_q\rangle
P^{qs}_{or}(D^{-1})_p^o,
\end{eqnarray}
where $D$ and $P$ are Hermitialized one{\color{black}-} (1RDM) and two{\color{black}-} (2RDM) particle reduced density matrices
defined in \cite{sato2018communication}, and $W^r_s$
is the mean-field operator \cite{sato2013time}.
The matrix element $X^q_p$ includes orbital rotations among various subspaces.
Non-redundant orbital rotations {\color{black} are} determined by
$i\left(\delta^a_bD^j_i-D^a_b\delta^j_i\right)X^b_j = \label{eq:occ_eom_hp}
F^a_pD^p_i - D^a_pF^{i*}_p
-\frac{i}{8} \dot \tau_{ijk}^{abc}\lambda_{bc}^{jk} - \frac{i}{8} \tau_{ijk}^{abc}\dot\lambda_{bc}^{jk}.$
Redundant orbital rotations $\{X^{i^\prime}_{j^\prime}\}$,
$\{X^i_j\}$, and $\{X^a_b\}$ can be arbitrary
antiHermitian matrix elements.
The general expressions for the RDMs are {\color{black} the} same as in the TD-OCCDT(4) method \cite{pathak2021time}.
\begin{table}[!t]
\caption{\label{tab:gs} Comparison of the ground state energy of BH (r$_e$=2.4 bohr) molecule in DZP basis$^a$.}
\begin{center}
\begin{tabular}{lccl}
\hline
\hline
Method& This work & Reference &\\
\hline
OCCD$^b$ &$-$25.225\,591\,67&$-$25.225\,592& \cite{bozkaya2012symmetric}\\
OCCD(T)$^b$& $-$25.226\,913\,29&$-$25.226\,913&
\cite{bozkaya2012symmetric}\\
OCCD$^c$ &$-$25.178\,285\,70&$-$25.178\,286&
\cite{krylov1998size}\\
OCCD(T)$^c$&$-$25.178\,301\,00&&\\
\hline
\hline
\end{tabular}
\end{center}
{(a) Gaussian09 program \cite{gaussian09} is used to generate the required one-electron, two-electron, and overlap integrals, required for the imaginary time propagation of EOMs in the orthonormalized gaussian basis. A convergence cut-off of 10$^{-15}$ Hartree of energy difference is chosen in subsequent time steps.
(b) Six electrons correlated within the full basis set. (c) Six electrons correlated within the six optimized active orbitals.}
\end{table}

\section{Numerical Results and Discussion}\label{sec3}
Our numerical implementation has an interface with the Gaussian09 program \cite{gaussian09} for checking ground state energy with the standard Gaussian basis results.
We study BH molecule with double-$\zeta$ plus polarization (DZP).
We have reported ground state energy computed by propagating in the imaginary time for OCCD and OCCD(T) methods in Table \ref{tab:gs} and compare{\color{black}d those} with the {\color{black} optimized double and asymmetric triple excitation corrections for the orbital-optimized doubles method of Bozkaya {\it et al.,} \cite{bozkaya2012symmetric}. We also compare our OCCD ground state energy result with Krylov {\it et al.,}\cite{krylov1998size} within the chosen active space of six electrons correlated among the six optimized active
orbitals}.
We obtained a perfect agreement for all available values.

We have used a spherical-finite-element-discrete-variable representation (FEDVR) basis for representing orbital functions, \cite{sato2016time, orimo2018implementation}
$\chi_{klm}(r, \theta, \psi)=\frac{1}{r}f_k(r)Y_{lm}(\theta, \phi)$ 
where $Y_{lm}$ and $f_k(r)$ are spherical harmonics and the normalized radial-FEDVR basis function, respectively.
The expansion of the spherical harmonics continued up to the maximum angular momentum $L_{max}$, and the radial FEDVR basis supports the range of radial coordinate $0\leq r \leq R_{max}$,
with cos$^{1/4}$ mask function used as an absorbing boundary for avoiding unphysical reflection from the wall of the simulation box.
We have used $l_{\text{max}}=72$, and the FEDVR basis supporting the radial
coordinate $0 < r < 300$ using 78 finite elements each containing 25 DVR functions.
The absorbing boundary {\color{black} is} switched on at $r = 180$ in all our simulations.
The Fourth-order exponential Runge-Kutta method
\cite{exponential_integrator} is used to propagate the EOMs with 20000
time steps for each optical cycle. 
We run the simulations for a further 6000 time steps after the end of the pulse.
In all correlation calculations, eight electrons of $4s4p$ orbitals are considered as active and correlated among thirteen active orbitals.
We report simulation results computed using a three-cycle laser pulse with
a central wavelength of 800 nm 
having intensity 2$\times 10^{14}$ W/cm$^2$ and a period of $T=2\pi/\omega_0 \sim 2.67$ fs. 
\begin{figure}[!]
\centering
\includegraphics[width=0.7\linewidth]{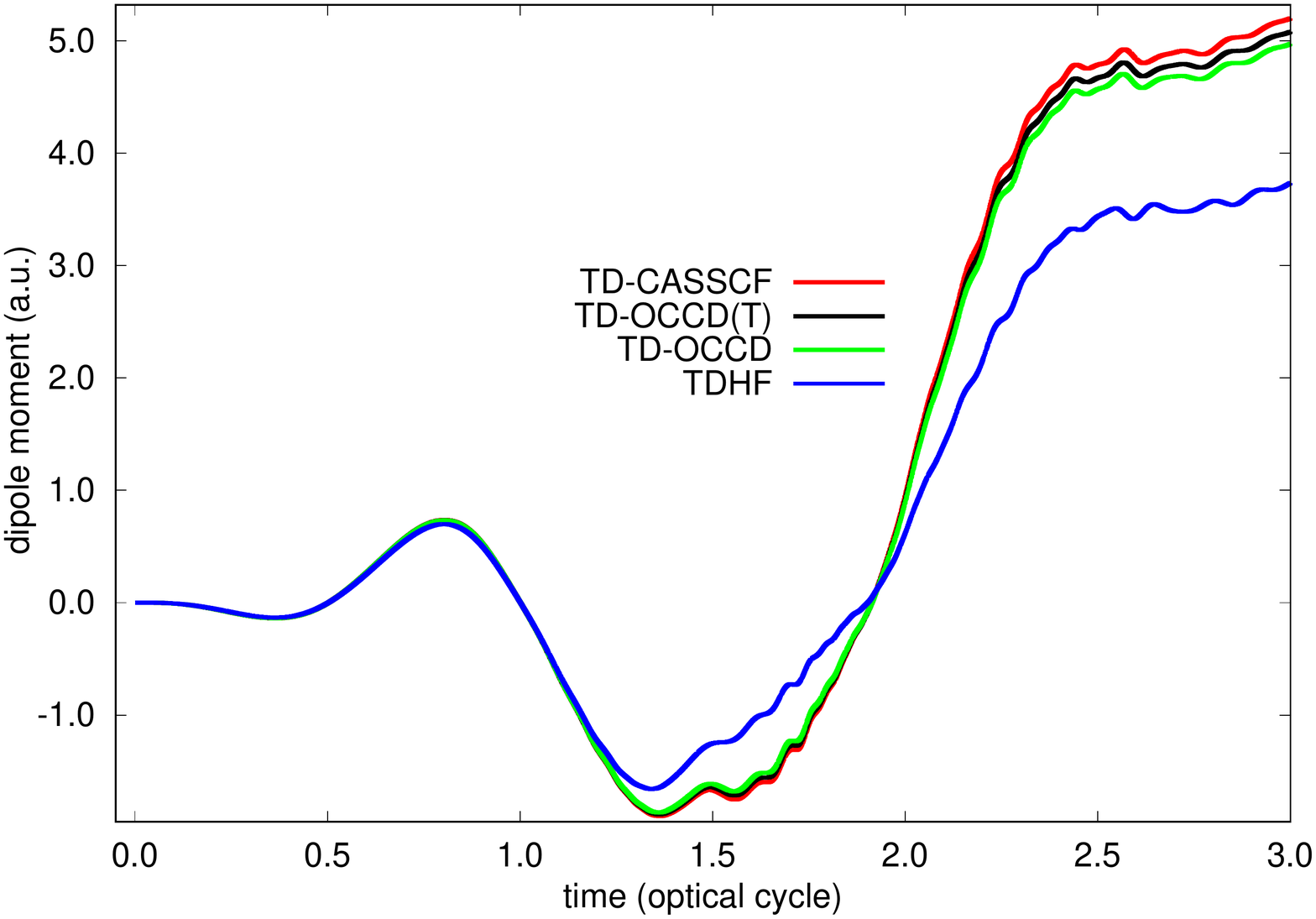}
\caption{\label{fig:kr_dipole}
{\color{black} Time evolution of dipole moment of Kr irradiated by
a laser pulse with a wavelength of 800 nm and a peak intensity of 
2$\times$10$^{14}$ W/cm$^2$ 
calculated with TDHF, TD-OCCD, TD-OCCD(T), and TD-CASSCF
methods.}}
\end{figure}

\begin{figure}[!t]
\centering
\includegraphics[width=0.7\linewidth]{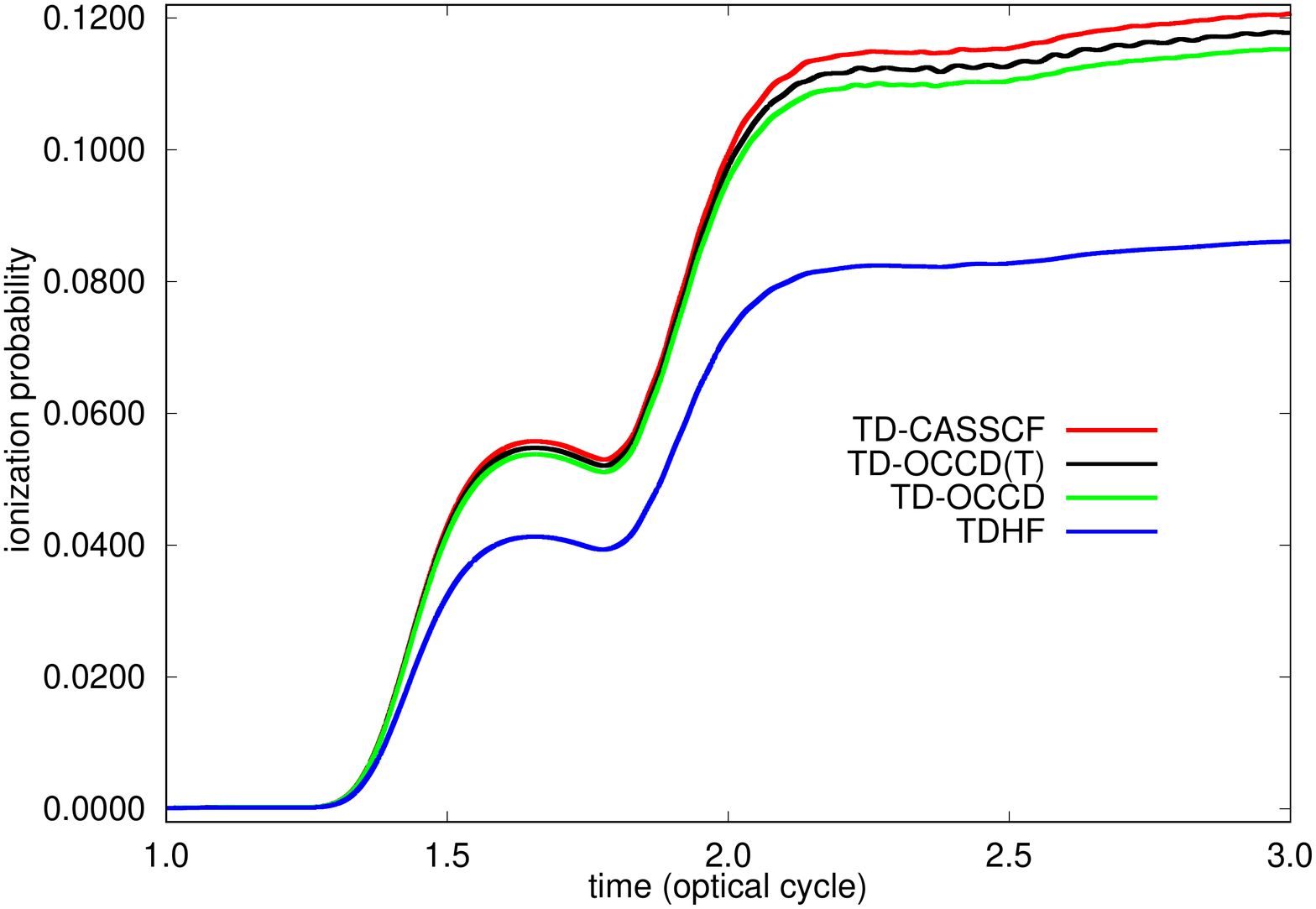}
\caption{\label{fig:kr_ionization}
{\color{black} Time evolution of single ionization probability of Kr irradiated by
a laser pulse with a wavelength of 800 nm and a peak intensity of
2$\times$10$^{14}$ W/cm$^2$
calculated with TDHF, TD-OCCD, TD-OCCD(T), and TD-CASSCF
methods.}}
\end{figure}

\begin{figure}[!t]
\centering
\includegraphics[width=0.7\linewidth]{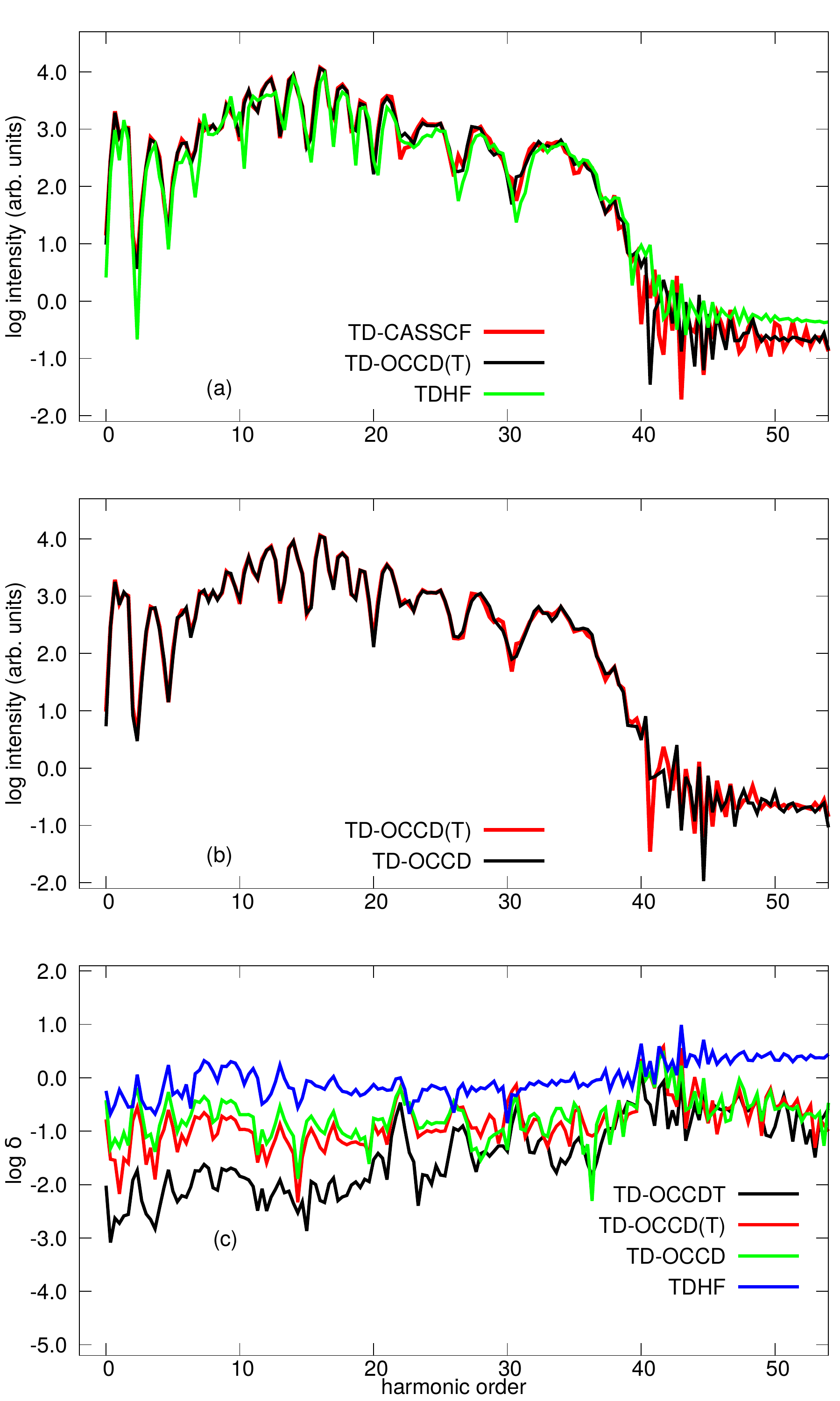}\\
\caption{\label{fig:krhhg2}
The HHG spectra (a) and the relative deviation (b) of the spectral amplitude from the TD-CASSCF spectrum from Kr irradiated by a laser pulse
with a wavelength of 800 nm and 
a peak intensity of 2$\times$10$^{14}$ W/cm$^2$ with various methods.}
\end{figure}

We report the time evolution of dipole moment of Kr in Fig.~\ref{fig:kr_dipole} and 
in Fig.~\ref{fig:kr_ionization} single electron ionization probability.
Time-dependent dipole moment is evaluated as a trace
$\langle \psi_p|\hat z
|\psi_q\rangle {\color{black}D^q_p}$
using 1RDMs. 
For the single electron ionization probability, 
we computed the probability of finding an electron outside a sphere of a radius of 20 a.u.
using RDMs defined in Refs.~\citenum{fabian2015, fabian2017, sato2018chapter}.
We compare the results of TD-CASSCF, TD-OCCD(T), TD-OCCD, and TDHF methods.
 %
We observe a substantial underestimation (both in Fig.~\ref{fig:kr_dipole}, and Fig.~\ref{fig:kr_ionization}) by the TDHF method due to the lack of correlation treatment. All correlation methods perform according to their 
ability to treat electron correlation. 
We also computed results using the TD-OCCDT method but not reported here since those results are not identifiable from the TD-CASSCF results within the graphical resolution.

Next, we report high-harmonic generation in Fig.~\ref{fig:krhhg2}.
It is calculated by squaring the modulus $I(\omega) = |a(\omega)|^2$ of the Fourier transform of the expectation value of the dipole acceleration with a modified Ehrenfest expression \cite{sato2016time}.
In panel (c) of Fig.~\ref{fig:krhhg2}, we plot the absolute relative deviation ($\delta(\omega)$, of the spectral amplitude $a(\omega)$ from the TD-CASSCF value for each method. 
All methods qualitatively predict similar HHG spectra with TDHF underestimates the spectral intensity. The relative deviation of results from TD-CASSCF ones follows the general trend TDHF$>$TD-OCCD$>$TD-OCCD(T)$>$TD-OCCDT, the same as what we observe for the time-dependent dipole moment and single ionization probability. We also simulated results with lower and higher intensity. However,
the trend remains the same.

Finally, we make a tally of computational costs for all the methods considered in this article.
All simulations performed using an Intel(R) Xeon(R) Gold 6230 central processing unit (CPU) with 40 processors with a clock speed of 2.10 GHz, and report total simulations time in Table~\ref{tab:timing_electron}.
Further, we report a reduction in the computational cost for various TD-OCC methods relative to the TD-CASSCF. 
We see a massive 63\% cost reduction for the TD-OCCD(T) method, which is larger than for the TD-OCCDT method (58\%), and a minimal increase from the TD-OCCD method.

\section{Concluding Remarks}\label{sec4}
We have reported {\color{black} the} formulation and implementation of the TD-OCCD(T) method.
As the first application, we employed this method to study laser-driven dynamics in Kr exposed to an intense near-infrared laser pulse.
We observe a 63 \% cost reduction in comparison to the TD-CASSCF method without losing much accuracy.
Therefore, we conclude that TD-OCCD(T) method will certainly be beneficial in exploring highly accurate {\it ab initio} simulations of electron dynamics in larger chemical systems.

\begin{table}[!t]
\caption{\label{tab:timing_electron} \color{black} Comparison of the total simulation time$^\text{a}$ (in min) spent for TD-CASSCF, TD-OCCDT, TDCCD(T), and TD-OCCD methods}
\begin{center}
\begin{tabular}{lrr}
\hline
\hline
Method & \multicolumn{1}{c}{Time (min)} & cost reduction (\%)\\
\hline
TD-CASSCF &\,47303 &\,\dots \\
TD-OCCDT &\,19697 &\, 58\\
TD-OCCD(T) &\,17504 &\,63\\ 
TD-OCCD&\,17494&\,63\\
\hline
\hline
\end{tabular}
\end{center}
{(a) Time spent for the simulation of Kr atom for 66000 time steps ($0 \leq t \leq 3.3T$) of a real-time simulation ($I_0=2\times 10^{14}$ W/cm$^{2}$ and $\lambda=800$ nm.),
 using an Intel(R) Xeon(R)  Gold 6230 CPU with 40 processors having a clock speed of 2.10GHz.}                                
\end{table}

\section*{Acknowledgement}
This research was supported in part by a Grant-in-Aid for
Scientific Research (Grants No. JP18H03891 and No. JP19H00869) from the Ministry of Education, Culture,
Sports, Science and Technology (MEXT) of Japan. 
This research was also partially supported by JST COI (Grant No.~JPMJCE1313), JST CREST (Grant No.~JPMJCR15N1),
and by MEXT Quantum Leap Flagship Program (MEXT Q-LEAP) Grant Number JPMXS0118067246.
\bibliography{2_references}

\end{document}